\documentclass[a4paper,12pt]{article}

\usepackage{epsfig}
\usepackage{amsmath} 
\usepackage{amssymb}
\renewcommand{\Re}{\mathfrak{Re}}
\renewcommand{\Im}{\mathfrak{Im}}

\newcommand{\nonu}{\nonumber}
\newcommand{\non}{\nonumber \\}
\newcommand{\mbf}{\mathbf}
\newcommand{\lp}[1][(]{\left#1}
\newcommand{\rp}[1][)]{\right#1}
\newcommand{\up}{\textsuperscript}
\newcommand{\ket}[1]{| {\bf #1} \rangle}
\newcommand{\bra}[1]{\langle {\bf #1} |}
\newcommand{\sumpl}[3][]{\sum\limits_{l_{#1}=0}^{+\infty} \, (2l+1)
P_{l_{#1}}(\cos(\widehat{{\bf #2},{\bf #3}}))}
\newcommand{\summ}[1][]{\sum\limits_{m_{#1}=-\infty}^{+\infty}}
\newcommand{\hlp}[2][l]{h_{#1}^{(+)}(#2)}
\newcommand{\jl}[2][l]{{j_{#1}}(#2)}
\newcommand{\hm}[2][m]{H_{#1}^{(1)}(#2)}
\newcommand{\jm}[2][m]{{J_{#1}}(#2)}
\newcommand{\tk}[3][]{\langle {#2} | {\bf t}_K | {#3} {\rangle}_{#1}}
\newcommand{\ti}[3][]{\langle {\bf #2} | {\bf t}_{#1}(E) | {\bf #3} \rangle}
\newcommand{\T}{\ensuremath{\mathcal{T}}}
\newcommand{\go}[1][+]{{\bf G}_0^{(#1)}(E)}
\newcommand{\hb}{\frac{\hbar^2}{2m}}
\newcommand{\eps}{\epsilon}
\newcommand{\ord}{\mathcal{O}}
\newcommand{\dimp}[2][3]{\frac{d\mathbf{#2}}{(2\pi)^{#1}}}
\newcommand{\tm}[1][ ]{${\bf t}$-matrix#1}
\newcommand{\tme}[1][ ]{${\bf t}$-matrix element#1}
\newcommand{\tms}[1][ ]{${\bf t}$-matrices#1}
\newcommand{\mfp}[1][ ]{mean free path#1}
\newcommand{\Bmfp}[1][ ]{Boltzmann mean free path#1}
\newcommand{\ie}[1][ ]{\emph{i.e.#1}}
\newcommand{\appr}[1][ ]{approximation#1}
\newcommand{\pro}[1][ ]{propagator#1}
\newcommand{\se}[1][ ]{self-energy#1}
\newcommand{\pl}[1][ ]{point-like#1}

\newcommand{\JPCM}{{J. Phys.: Condens. Matter}}

\newcommand{\PLA}{{Phys. Lett.} \textbf{A}}

\newcommand{\RMP}{{Rev. Mod. Phys.}}

\title{Corrections to the Boltzmann mean free path in disordered
  systems with finite size scatterers}
\author{S. Correia\thanks{correia@lpt1.u-strasbg.fr}\\
{\small 
Laboratoire de Physique Th\'eorique}\\
{\small
3 rue de l'Universit\'e}\\
{\small
 F-67084 Strasbourg Cedex}}

\date{\today}

\begin{document}

\maketitle

\abstract{The \mfp is an essential characteristic length in disordered
  systems.  In microscopic calculations, it is usually approximated by
  the classical value of the elastic \mfp[].  It corresponds to the
  Boltzmann \mfp when only isotropic scattering is considered, but it
  is different for anisotropic scattering.  In
  this paper, we work out the corrections to the so called \Bmfp due
  to multiple scattering effects on finite size scatterers, in the
  $s$-wave approximation, \ie when the elastic \mfp is equivalent to
  the \Bmfp[].  The main result is the expression for the \mfp
  expanded in powers of the perturbative parameter given by the
  scatterer density.}

\section{Introduction}
The mean free path $l$ is an important parameter which allows to
characterize the different regimes in disordered systems
\cite{ir,sheng}.  When the disorder is strong, there appears a
localized phase for $Kl\ll 1$, where $K$ is the wave number, whereas
the diffusive phase is characterized by $Kl\gg 1$.  The strength of
disorder is thus measured by the parameter $1/Kl$.

In most theoretical approaches, the randomly distributed impurities
are often approximated by point-like potentials
\cite{devries,tiggelen,seba}.  Although this type of potential greatly
simplifies the analytical calculations, it has been found that this
approximation can lead to incorrect results, in particular as far as
causality violation is concerned \cite{pell}.  Scattering theory with
finite size potentials does not encounter any problem with causality
\cite{joa}.  Therefore, it is important to use finite size potentials
in order to describe multiple scattering effects.  Effects due to the
finiteness of the potential size are generally neglected even when the
description is in terms of scattering \tms which describe finite size
potentials \cite{mahan,obb}.  As we shall show, contributions coming from the
momentum dependence of the \tme[s] and from spatial correlations can
be neglected.  Nevertheless, one has to take care of the way 
 corrections due to multiple scattering are calculated.  It is indeed shown
in this paper that the range of the potential plays an important role
in the calculation of corrections to the Boltzmann mean free path in
the diffusive regime, where the scatterer density is low.  Finally,
the \mfp is expanded in powers of the scatterer density up to the
first subleading order in 2 and 3 space dimensions.

\section{Formalism}
We consider the propagation of an electron in a disordered system
which is described by the following Hamiltonian.

  \begin{align}
    H=&-\hb\Delta+\mbf V,
  \end{align}
  where the potential $\mbf V=\sum_{i=1}^N\mbf v_i$ is the sum of the
  $N$ randomly distributed individual potentials.  The simplest finite
  range interaction, which one can imagine, is the hard sphere
  potential defined by
  \begin{align}
    v_i(\mbf r)=v(|\mbf r-\mbf R_i|)=\begin{cases}
      \infty & \text{for }|\mbf r-\mbf R_i| \le a, \\
      0 & \text{for } |\mbf r-\mbf R_i| > a.
      \end{cases}
 \end{align}
 Like the point-like scatterer given by a delta function, this type of
 scatterer has no internal structure (the wave function vanishes
 inside the potential), but it has a finite size given by the range of
 the potential~$a$.  The electron propagator can be cast in the form
\begin{align}\label{eq:2}
  {G}=&G_0+G_0 \lp{\sum_{i=1}^N\mbf v_i}\rp {G} =G_0+G_0 {\T} G_0,
\end{align}
where $G_0$ is the free \pro and $\T$ is the global scattering matrix
of the system.  This $\T$-matrix is given in terms of the individual
scattering $\mbf t$-matrices by the Watson series~\cite{watson}
\begin{align}\label{watson}
  {\cal T} = & {\sum_{i=1}^N {\bf t}_i} +\sum_{i,j\neq i} {\bf t}_iG_0
  {\bf t}_j +\sum_{\scriptsize\begin{array}{c}{i,j\neq i,}\\{k\neq
        j}\end{array}} {\bf t}_iG_0 {\bf t}_jG_0 {\bf t}_k+ \cdots
\end{align}
Each term of this operator series represents a sequence of multiple
scattering.  The insertion of the closure relation
$\int\dimp[3]{k}\ket{k}\bra{k}=1$ between the operators introduces
off-shell \tme[s].  The hard sphere off-shell \tme[s], $\tk{\mbf
  k'}{\mbf k}$ with $k\neq k'\neq \sqrt{2mE}$, can be derived from the
off-shell barrier potential of finite height~\cite{schick} by
extrapolation of its height to infinity.  Their expression is given in
appendix~A.  The analytic expression, of each term of the electron
propagator in Fourier space, $G(E,\mbf k)$, given by~(\ref{eq:2}), is
obtained by inserting the momentum closure relation in the scattering
series~(\ref{watson}).  The momentum integrations are then performed
with the use of expressions~\eqref{eq:tmdiskdur} and~\eqref{hard_off}
for the \tme[s] in 2 and 3 dimensions and the
integrals~(\ref{eq:integrale}) and~(\ref{integrale3d}) of appendix~B.
Throughout the calculation, one has to take care of the fact that the
scatterers do not overlap.  This gives rise to correlations between
scatterers which have to be accounted for when taking the ensemble
average of the propagator.

In contrast to the point-like scatterers case where there is no
spatial correlation, the ensemble average, noted with a bar, of a
quantity $Q$ reads
\begin{align}\label{eq:vraimoy}
  \overline{Q}=\frac{1}{\mathcal{Z}} \int \prod_{i=1}^{N} d{\bf R}_i
  \,\prod_{j>i} \Theta(R_{ij}-2a)\, Q(\{{\bf R}_i\}_{i=1,\cdots,n}),
\end{align}
where $R_{ij}\equiv |{\bf R}_{ij}|\equiv |{\bf R}_{i}-{\bf R}_{j}|$
and
$$
\mathcal{Z}= \int \prod_{i=1}^{N} d{\bf R}_i \,\prod_{j>i}
\Theta(R_{ij}-2a)
$$
is a normalization constant, the same as the spatial term of the
partition function of a 3D hard sphere or 2D hard disk gas.  The
Heaviside step function $\Theta$ accounts for the spatial
correlations.

Using results from the virial expansion, especially the Kirkwood
superposition approximation \cite{hill2}, the spatial correlation
function can be written as a product of two-body correlation
functions~$g$.
\begin{equation}
\frac{1}{\mathcal{Z}}\prod_{j>i}
\Theta(R_{ij}-2a)=\frac{1}{V^n}\prod_{j>i} g(R_{ij}). 
\end{equation}
Then, the average of $Q$ can be approximated by
\begin{equation}
  \overline{Q}= \frac{1}{V^n} \int
  \prod_{i=1}^{n} d{\bf R}_{i}\prod_{j>i} g(R_{ij}) Q(\{{\bf
    R}_i\}_{i=1..n}),
\end{equation}
where the two-body correlation function $g$ is expanded in powers of
the scatterer density $\rho$
\begin{align}\label{eq:viriel-g}
  g(R_{ij})=g_0(R_{ij})+\rho g_1(R_{ij})+\rho^2g_2(R_{ij})+\cdots,
\end{align}
$g_0$, $g_1$, $g_2$\ldots being derived from the virial expansion
\cite{hill2}.  In the case of hard spheres,
$g_0(R_{ij})=\Theta(R_{ij}-2a)$ describes the fact that two hard
spheres at position $\mbf R_i$ and $\mbf R_j$ cannot overlap, and the
other terms $g_1$, $g_2$\ldots take into account the average effect
coming from the presence of other scatterers in the system.

In the case of point-like scatterers $g_0(R_{ij})=1$ and $g_{i\ge
  1}(R_{ij})=0$.  Notice that using this approximation, \ie neglecting
the spatial correlations, for the calculation of
$\overline{\bra{k'}\mbf t_i \go \mbf t_j \go \mbf t_i\ket k }$ with
the expressions~\eqref{hard_off} and~\eqref{eq:tmdiskdur} leads to a
divergence instead of the correct result \cite{correia}.  This
divergence is due to the convergence conditions of
integrals~(\ref{eq:integrale}) and~(\ref{integrale3d}).  It shows that
the use of off-shell \tme[s] is restricted to non-overlapping
scatterers.  This can be understood by observing that the existence of
off-shell \tme[s] is related to the finiteness of the size of the
scatterer.  Therefore it makes no sense to use these off-shell \tme[s]
in a \pl scatterer \appr[].

In order to calculate the average of the \pro[] analytically, one
needs an additional \appr[], which neglects the correlations between
non successive scatterers which appear in the scattering
series~(\ref{watson}).  Then
\begin{equation}\label{eq:Vexclu}
  \overline{Q}= \frac{1}{V^n} \int d{\bf R}_n \int
  \prod_{i=1}^{n-1} d{\bf R}_{i,i+1} g(R_{i,i+1}) Q(\{{\bf
    R}_i\}_{i=1..n}).
\end{equation}
With the help of this \appr[], each sequence of the Watson
series~\eqref{watson} containing only distinct scatterers can be
calculated.  The summation of this class of terms is usually referred
to as \emph{independent scatterer \appr[]} \cite{rossum}.  It gives
the first order contribution in density of the self-energy.  The
\appr[]~\eqref{eq:Vexclu} is justified as long as the self-energy
$\Sigma$ of the average \pro

$$\bar{G}(E,\mbf k)=G_0(E,\mbf k)+G_0(E,\mbf k)\Sigma(E,\mbf k)
\bar{G}(E,\mbf k)$$
is expanded up to order $\rho^2$,
$$\Sigma(E,\mbf k)=\underbrace{\Sigma^{(1)}(E,\mbf k)}_{\ord(\rho)}+
\underbrace{\Sigma^{(2)}(E,\mbf k)}_{\ord(\rho^2)}+\cdots$$
Indeed, corrections coming from the overlapping of two non-successive
scatterers are at least of the order $\rho^3$, because they correspond
to scattering sequences containing at least three scatterers.

In the $s$-wave \appr[], where $Ka\ll 1$, the first order term of the
\se is simply given by the off-shell \tme
\begin{equation}
  \label{eq:sig1}
  \Sigma^{(1)}(E,\mbf k)\approx \rho\ti{k}{k}.
\end{equation}
Notice that this expression is the one obtained in the usual
independent scatterer \appr[].  Its imaginary part is related to the
individual scattering cross section via the optical theorem and leads
to the \Bmfp $l_0$ \cite{mahan},
\begin{equation}
  \label{eq:Bmfp}
  l_0=-\frac{K}{\Im \Sigma^{(1)}}\approx \begin{cases}
    \displaystyle\frac{K\ln^2 Ka}{\rho\pi^2} & \mbox{ in 2D,} \\
    \displaystyle\frac{1}{4\pi\rho a^2} & \mbox{ in 3D,} \end{cases}
\end{equation}
where the \se is taken on the energy shell, $\Sigma^{(1)}\equiv
\Sigma^{(1)}(E,\mbf k)$ with $|\mbf k|=K=\frac{\sqrt{2mE}}{\hbar}$.

The second term of the \se contains the contribution coming from the
resummation of the infinite series of all multiple scattering terms
involving two distinct scatterers and a further term coming from the
excluded volume.
It appears that this latter term is negligible in comparison with the
other terms of order $\rho^2$, because of order $a^d$ where $d$ is the
spatial dimension.  The two-body scattering term is calculated in
appendix C and reads
\begin{equation}\label{eq:sigma2-2d}
  \Sigma^{(2)}\approx \hb
    \lp\frac{\pi\rho}{K}\rp^2\left\{\frac{2}{\ln^3 K'a} +\frac{3\pi
        i}{\ln^4 K'a}\right\}
\end{equation}
where $K'=Ke^\gamma/2$ and $\gamma\simeq 0.5772$ is the Euler constant
in 2D, and
\begin{equation}\label{eq:sigma2-3d}
   \Sigma^{(2)}\approx \hb \frac{(4\pi)^2a^3}{K} \rho^2
   \left\{\frac{i}{2}+Ka \left(\ln 
        Ka+3\ln2+\ln3 +\gamma-3-\frac{i\pi}2\right)\right\}
\end{equation}
in 3D.

\section{Corrections to the Boltzmann mean free path}
The \mfp $l$ is defined as the decreasing rate of the average \pro
$$
\bar G(E,\mbf r-\mbf r')\propto e^{-\frac{|\mbf r-\mbf r'|}{2l}}.
$$

In order to compare with the case of point-like scatterers
\cite{tiggelen}, we consider the deviation of the \mfp with respect to
the \Bmfp $l_0$.
\begin{align}\label{eq:3}
  \frac l {l_0}=& 1-
  \lp\frac{\Re{\Sigma}^{(1)}}{2K^2}-\frac{\Re{\Sigma}^{(1)\prime}}{K}+
  \frac{\Im\Sigma^{(2)}}{\Im\Sigma^{(1)}}\rp+\ord(\rho^2),
\end{align}
where $\Sigma^{(1)\prime}\equiv \frac{d}{dk}\left. \Sigma^{(1)}(E,\mbf
  k)\right|_{k=K}$.  This formula is easily derived from the
definition of the \mfp[],
\begin{equation}
  \label{eq:mfpdef}
  l=\frac{1}{2\Im \tilde{K}}, \mbox{ where } \tilde{K}^2=
  K^2-(\Sigma^{(1)}+\Sigma^{(2)}). 
\end{equation}
For \pl scatterers, the term $\Sigma^{(1)\prime}$ vanishes, there is
no off-shell contribution.\\

In 2 dimensions, for low density $\rho$ and low energy $Ka\ll1$, one
obtains
\begin{equation}\label{eq:1}
    \frac{l}{l_0}\approx 1+\frac{\pi\rho}{K^2}\lp
    \frac{1}{\ln Ka}+\frac{3}{\ln^2 Ka}\rp.
\end{equation}  
The excluded volume term appears to be of the same order in $Ka$ as
the off-shell term $\Sigma^{(1)\prime}$, \ie $\ord\lp(Ka)^2\rp$.
These two terms are also of the same order in $Ka$ as the first
partial wave term which is the term for $m=\pm 1$ in
expression~(\ref{eq:tmdiskdur}).  They are therefore neglected here.

The leading correction term in~(\ref{eq:1}) is given by the on-shell
\tm $\Sigma^{(1)}$ and the subleading correction term comes from the
first term of the two-body series.  These corrections lead to a
decrease of the \mfp with respect to the Boltzmann value.  Neglecting
the subleading corrections in $1/\ln^2Ka$, expression~(\ref{eq:1}) can
be written in terms of $1/Kl_0$ and reads
\begin{equation}\label{eq:11}
    \frac{l}{l_0}\approx 1+
    \frac{\ln Ka}{\pi}\frac{1}{Kl_0}.
\end{equation}
This expression shows that the corrections are not only written in terms of $1/Kl_0$ but that there appears a logarithmic correction involving the size of the scatterer. \\

In 3 dimensions, the excluded volume and the off-shell corrections are
also negligible for the same reasons.  Contrary to the two dimensional
case, the leading term is not given by the on-shell \tm[]
$\Sigma^{(1)}$.  This latter is compensated by the first term of the
two-body series $\Sigma^{(2)}$, given by~(\ref{eq:I-iji}).  It comes
out that the leading term of the corrections to the \Bmfp is given by
the second term~(\ref{eq:I-ijij}) of the two-body series.  As in 2
dimensions, the \mfp is smaller than the Boltzmann expression.

\begin{equation}\label{eq:4}
 \frac l {l_0}\approx 1-\rho\lp\frac{\pi^2
        a^2}{K}+\ord(a^3)\rp.
\end{equation}

The expression~(\ref{eq:3}) for the deviation of the \mfp already
exists for \pl scatterers in 3 dimensions \cite{tiggelen}.  Instead of
being of the form $\rho a^2/K$ as obtained in~(\ref{eq:4}), the
corrections to the \Bmfp for scalar scatterers are proportional to
$\rho/K^3$.  This shows that the behavior of the scalar \pl scatterers
is very different from that of finite size scatterers.  It is much
more sensitive to small wave numbers.  For finite size potentials, the
effect of two-body scattering is much less important than it is in the
case of \pl potentials.  Written in terms of the disorder strength
$1/Kl_0$, expression~(\ref{eq:4}) reads
\begin{equation}\label{eq:44}
 \frac l {l_0}\approx 1-\frac{\pi}{4}\frac{1}{Kl_0}.
\end{equation}
For the scalar point scatterers at resonance, a similar expression is
found except that the numerical factor is not $\pi/4$ but 0.375 (see
eq.~(3.14a) from~\cite{tiggelen}).  The effect of the disorder is more important for finite size scatterers than for \pl scatterers.

\section{Conclusion}
In this paper we worked out the corrections to the \Bmfp of infinite
disordered systems composed of hard disk scatterers in two dimensions
and hard sphere scatterers in three dimensions.  Calculations with
finite size scatterers present several differences compared with
calculations with \pl scatterers.  One is led to use off-shell
\tme[s].  The excluded volume has to be taken into account, yielding
spatial correlations when averaging over disorder.  These two effects
appear to be of the same order in $Ka$ and are negligible with respect
to a more important effect due to multiple scattering.  They become
important when one is interested in non-isotropic scattering, \ie when
more than one partial wave has to be taken into account.

To the knowledge of the author, the result obtained in two dimensions
is new and has not been evaluated even for \pl potentials.  The
corrections to the \mfp cannot be given only in terms of $1/Kl_0$, indeed
they contain a logarithmic prefactor depending on the size of the
scatterer.  In three dimensions, the comparison of our result with the
result obtained for \pl potentials shows a discrepancy in the
numerical factor in front of the disorder strength $1/Kl_0$.  The
decrease of the \mfp is more important for finite size scatterers than
for \pl potentials.\\

\textbf{Acknowledgment}

I would like to thank D.~Boos\'e, J.-M.~Luck and J.~Richert for useful
discussions.

\setcounter{section}{1}
\section*{Appendix \Alph{section} : Off-shell \tm}
\label{sec:append-alphs-}

Following \cite{schick},  the off-shell scattering \tm
of a quantum particle with energy $E=\frac{\hbar^2K^2}{2m}$ takes the form
\begin{align}\label{eq:tmdiskdur}
  \ti{\bf{k'}}{\bf{k}}=& \dfrac{\hbar^2}{2mV}2 \pi a \summ e^{i m
    (\theta_{k'} - \theta_k)} \tag{\Alph{section}.1} \\ 
  &\left\{\dfrac{k^2-K^2}{k^{\prime
        2}-k^2}(k \jm{k'a} \jm[m-1]{ka}
    - k' \jm[m-1]{k'a} \jm{ka}) \right.\non
  &\left. +k \jm{k'a} \jm[m-1]{ka} - K \jm{ka} \jm{k'a}
    \dfrac{\hm[m-1]{Ka}}{\hm{Ka}}\right\}\nonu
\end{align}
in 2 dimensions and
\begin{align}\label{hard_off}
  \ti{{\bf k'}}{{\bf k}} =& \dfrac{2\pi \hbar^2 }{mV} a^2
  \sumpl{k'}{k}\tag{\Alph{section}.2}  \\ & \times\Bigg\{\dfrac{k^2-K^2}{k^{\prime
      2}-k^2}\bigg[k\jl{k'a}\jl[l-1]{ka}
  -k'\jl[l-1]{k'a}\jl{ka}\bigg] \non
  & + k\jl{k'a}\jl[l-1]{ka} -
  K\jl{k'a}\jl{ka}\dfrac{\hlp[l-1]{Ka}}{\hlp{Ka}}\Bigg\}\nonu
\end{align}
in 3 dimensions. The $j_l$ and $h_l^{(+)}$ are the spherical Bessel
and Hankel functions and the $J_m$ and $H_m^{(1)}$ are the ordinary
Bessel and Hankel functions \cite{grad}.  In 3 dimensions, the angular
part is taken into account by the Legendre polynomial~$P_l$.

These expressions enter the calculation of each term of the
scattering series~(\ref{watson})  in the expression of the electron
propagator~(\ref{eq:2}).

\stepcounter{section}
\section*{Appendix \Alph{section} : Useful integrals}

In order to derive the expressions of each term of the scattering
series~(\ref{watson}) in Fourier space, one has to compute products of
propagators $G_0$ with \tms[].  The following integrals are the basic
integrals appearing in these products.  

The  integral used in the calculation of each scattering
sequence of the Watson series~(\ref{watson}) in 2 dimensions reads
\begin{align}\label{eq:integrale}
        \int_0^{+\infty}\frac{k^{d-1}dk}{K^2-k^2+i \eps}
        &\prod_{i=1}^n\jm[\mu_i]{ka_i}\jm{kR}\non
        =&-i \frac{\pi}{2}K^{d-2}
        \prod_{i=1}^n\jm[\mu_i]{Ka_i}\hm{KR}.\tag{\Alph{section}.1} 
\end{align}
It is valid under the following conditions
\begin{equation}\label{Cconv}
\left\{
\begin{array}{l}
        d +m+\sum_{i=1}^n \mu_i=2p   \qquad  p\in \mathbb{Z}\\
        R >\sum_{i=1}^n a_i \\
        d +\sum_{i=1}^n |\mu_i| > |m|. 
\end{array}
\right.\nonu
\end{equation}

In 3 dimensions, one gets
\begin{align}\label{integrale3d}
    \int_0^{+\infty} \frac{k^{d-1} dk}{K^2-k^2+i \eps}
    &\prod_{i=1}^n \jl[\mu_i]{ka_i}\jl{kR}\non
    =&-\frac{\pi}{2}K^{d-2} 
    \prod_{i=1}^n \jl[\mu_i]{Ka_i}\hlp{KR},\tag{\Alph{section}.2} 
\end{align}
if the conditions 
\begin{equation}\label{Cconv3d}
\left\{
\begin{array}{l}
        d +1 +l+\sum_{i=1}^n \mu_i=2p   \qquad  p\in \mathbb{Z}^+ \\
        R \ge \sum_{i=1}^n a_i \\
        d -1+\sum_{i=1}^n \mu_i > l
\end{array}
\right.\nonu
\end{equation}
are verified.

\stepcounter{section}
\section*{Appendix \Alph{section} : Two-body scattering}

The two-body scattering term consists of the summation of all the
multiple scattering terms with two distinct scatterers
\begin{equation}
  \label{eq:series}
  (iji)+(ijij)+(ijiji)+(ijijij)+\cdots
\end{equation}
where $(ij)$ is a symbolic notation for $\sum_{i,j\neq i} {\bf t}_iG_0
{\bf t}_j$.

In 2 dimensions, the average of the first term of this series is the
leading term and can be calculated exactly \cite{grad}.
\begin{align}
  \Sigma^{(2)} \approx &-8
  i\pi\rho^2\hb\left(\frac{\jm[0]{z}}{\hm[0]{z}}\right)^3
  \int_{2a}^{\infty}rdr\hm[0]{Kr}\hm[0]{Kr}\non
  =&16i\pi\rho^2a^2\hb\left(\frac{\jm[0]{z}}{\hm[0]{z}}\right)^3\left[(\hm[0]{2z})^2+
    (\hm[1]{2z})^2\right],\nonu
\end{align}
where $z=Ka$.  When $Ka\to 0$, expression~\eqref{eq:sigma2-2d} is
recovered.  The average of the other terms appearing in the
series~(\ref{eq:series}) can be evaluated by a saddle-point \appr of
$\int rdr \lp \hm[0]{Kr}\rp^n$.  They lead to an infinite series of
powers of $t=\frac{\jm[0]{z}}{\hm[0]{z}}\approx
\frac{1}{1+\frac{2i}{\pi}\ln \frac{z}{2}e^\gamma}$ for the \se[].
\begin{equation}
  \label{eq:se}
  \Sigma^{(2)}\approx -\frac{8i\pi}{K^2}\rho^2\hb t \sum_{n\ge2} (-)^n a_{n}t^n
\end{equation}
with $a_n\approx -\frac2{e^{2\gamma}}\lp\frac i\pi\rp^nn!$. 

In 3 dimensions, the geometrical series is first resummed before
averaging.  Then the order $\rho^2$ term of the self-energy is given
by the average of this resummation and reads
\begin{multline}\label{eq:8}
  \Sigma^{(2)}= \hb \frac{(4\pi)^2}{K} \rho^2
  \left(\frac{\jl[0]{z}}{\hlp[0]{z}}\right)^3 \\
  \times\int_{2a}^\infty \frac{r^2dr \hlp[0]{Kr}^2\left( 1-
      \frac{\jl[0]{z}}{\hlp[0]{z}} \jl[0]{Kr} \hlp[0]{Kr}\right)
    }{1-\left[\frac{\jl[0]{z}}{\hlp[0]{z}} \hlp[0]{Kr}
    \right]^2},\nonu
\end{multline}
The integral is decomposed into three parts, noted $I^{(1)}$,
$I^{(2)}$, and $I^{(3)}$.  The first contribution represents the
scattering sequence $(iji)$.
   \begin{equation}
     \label{eq:I-iji}
     I^{(1)}\equiv \frac{1}{z^2}\int_2^\infty dx \;e^{2izx}=
     \frac{ie^{4izx}}{2z^3}\approx\frac{i}{2z^3}-\frac{2}{z^2}.\tag{\Alph{section}.1} 
   \end{equation}
   The second contribution is the scattering sequence $(ijij)$.
    \begin{align}
      \label{eq:I-ijij}
      I^{(2)}\equiv& -\frac{\sin ze^{-iz}}{z^4}\int_2^\infty
      \frac{dx}{x^2}e^{3izx}\sin zx\non \approx&\frac{1}{z^2}\left(\ln
        z+4\ln2+\gamma-1-\frac{i\pi}2\right).\tag{\Alph{section}.2}
    \end{align}
    In the limit $z\to 0$, the remainder of the series can be
    evaluated by taking $z=0$.  This gives
\begin{equation}
  \label{eq:I-3}
  I^{(3)}=\frac{1}{z^2}
  \int_2^\infty\frac{dx}{x(x+1)}=\frac{1}{z^2}\ln\frac32.\nonu
\end{equation}
Combining these three expressions, one obtains
expression~\eqref{eq:sigma2-3d} for the \se[].

\end{document}